\documentclass{article}%
\usepackage{amsmath}
\usepackage{amsfonts}
\usepackage{amssymb}
\usepackage{graphicx}%
\begin{document}

\title{Upper limit on the transition temperature for non-ideal Bose gases}
\author{Wu-Sheng Dai$^{1,2}$ \thanks{Email: daiwusheng@tju.edu.cn} and Mi Xie$^{1,2}$
\thanks{Email: xiemi@tju.edu.cn}\\{\footnotesize $^{1}$ Department of Physics, Tianjin University, Tianjin
300072, P. R. China }\\{\footnotesize $^{2}$ LiuHui Center for Applied Mathematics, Nankai University
\& Tianjin University,} {\footnotesize Tianjin 300072, P. R. China} }
\date{}
\maketitle

\begin{abstract}
In this paper we show that for a non-ideal Bose gas there exists an upper
limit on the transition temperature above which Bose-Einstein condensation
cannot occur regardless of the pressure applied. Such upper limits for some
realistic Bose gases are estimated. This result implies that there may also
exist an upper limit on the transition temperature of superconductors.

\end{abstract}

%PACS numbers: 05.70.Fh; 03.75.Hh; 05.30.Jp

%Keywords: non-ideal Bose gas; Bose-Einstein condensation; transition temperature

\newpage

Bose-Einstein condensation is the first purely statistically derived example
of a phase transition. The problem of Bose-Einstein condensation in ideal and
non-ideal Bose gases has interested physicists both from practical viewpoint
and academic viewpoint for many years \cite{Griffin,Ziff}. Especially, in
recent years Bose-Einstein condensation in the dilute repulsive Bose gas has
attracted wide attention \cite{hsg}. In this paper, we will show that for a
non-ideal Bose gas there exists an upper limit on the transition temperature,
above which the condensation cannot occur regardless of the pressure applied,
though there is no upper limit on the transition temperature for an ideal Bose
gas. For an ideal Bose gas the condensation can occur in any temperature so
long as the pressure is large enough. In this paper, we will place an upper
limit on the transition temperature for Bose-Einstein condensation. We also
estimate such upper limits for some realistic Bose gases which are of course
non-ideal gases.

The reason why transitions may occur in Bose gases is that there exist
exchange interactions among the particles, and the effect of exchange
interactions is attractive; in other words, a Bose gas will exhibit attractive
forces, a pure quantum effect, between its gas molecules. The competition
between the exchange interaction and the thermal motion of molecules
determines the state of the system: when the exchange interaction dominates,
Bose-Einstein condensation takes place. The strength of the exchange
interaction, roughly speaking, is determined by the overlap among the
wavefunctions of the gas molecules. The overlap of the wavefunctions can be
roughly described by the magnitudes of the mean space between particles and
the mean thermal wavelength which measures the mean spatial extent of the
wavefunction of a particle. In the case of the ideal gas, there are no
interparticle interactions, so the separation between particles can take on
any small value. Therefore, the wavefunctions can overlap to any extent by
reducing the distance between particles no matter how high the temperature, or
no matter how short the wavelength, may be. As a result, ideal Bose gases can
display Bose-Einstein condensation at any temperature so long as the density
is high enough. That is to say, for ideal Bose gases there is no upper limit
on the transition temperature.

For non-ideal Bose gases, besides exchange interactions, there also exist
interparticle interactions. The interparticle interaction in a realistic gas
is attractive when the molecules are moderately far apart, and is repulsive
when they are close together. Especially, the repulsion will become so strong
that it can be idealized as infinite when the interparticle spacing is close
to a certain value. Consequently, there exists a lower limit on the
interparticle spacing due to the interparticle interaction, and thus the
overlap of the wavefunctions is limited. There must exist such a temperature
above which the corresponding mean thermal wavelength will be always shorter
than the interparticle spacing, and, then, the wavefunctions of the molecules
do not overlap. That is to say, for a non-ideal Bose gas, in contrast to the
ideal case, there is an upper limit on the transition temperature above which
Bose-Einstein condensation cannot occur regardless of the pressure applied. It
should be emphasized that, all realistic gases must have interparticle
interactions among their molecules, so for any realistic Bose gas there exists
an upper limit on the transition temperature.

The approach for obtaining the upper limit on the transition temperature for a
non-ideal Bose gas is based on the analogy between Bose-Einstein condensation
and the gas-liquid transition \cite{KU}. There exists an upper limit on the
transition temperature for gas-liquid transition, called critical temperature
(the word "critical temperature" here means the upper limit on the transition
temperature, not the phase transition temperature, though the transition
temperature is also called "critical temperature"). The existence of the upper
limit on the transition temperature in the gas-liquid transition is a result
of the existence of the interparticle interaction. The interparticle
interaction is attractive at a certain distance, which causes the gas to
condense into a liquid. Such an interaction, however, will become repulsive at
a very small distance, and consequently the potential energy has a minimum
value. Therefore, when the temperature is higher than a certain value --- the
upper limit on the transition temperature --- the gas cannot be liquefied
regardless of the pressure applied.

One approach for calculating the upper limit on the transition temperature for
the gas-liquid transition is first to obtain the van der Waals equation by
calculating the virial expansion of the equation of state for the non-ideal
gas, and then to calculate such an upper limit on the transition temperature
based on the van der Waals equation \cite{Greiner}. In principle, the result
obtained in such a way is valid only for the gas phase and cannot be used to
describe the gas-liquid transition since in gases the interaction between
particles is weak but in liquids the interaction between particles is strong.
However, after a modification which removes the unphysical part of the van der
Waals equation, which implies a negative compressibility, one can find a
result which can be used to describe the transition between the liquid and
gaseous states. Especially this result can be used to discuss the upper limit
on the transition temperature \cite{Pathria}.

In this paper, we use the hard-sphere Bose gas as a simplified model for
non-ideal Bose gases. This is because the essential reason for the existence
of an upper limit on the transition temperature is the existence of the
repulsive core in the interparticle interaction. The result obtained by the
hard-sphere model is also valid for more general interparticle interactions
since a particle that is spread out in space sees only an averaged effect of
the potential, and, thus often a complete knowledge of the detailed
interaction potential is not necessary for a satisfactory description
\cite{HYHYL}. Or, from the viewpoint of quantum mechanics, due to the low
collision energy of collisions between molecules in a Bose gas near the
transition temperature, one can use the $s$-wave approximation which is
shape-independent. The hard-sphere gas, as a simplified model, is of great
value for investigating the more general theory of non-ideal gases, especially
after Lee and Yang pioneered the study of interacting bosons \cite{LY58}; it
is often used as an efficient tool for studying Bose-Einstein condensation,
both for dilute atomic gases and liquid helium \cite{K}.

In the following we estimate the upper limit on the transition temperature for
the hard-sphere Bose gas. The equation of state for hard-sphere gases is
\cite{LY}%
\begin{align}
\frac{P}{kT} &  =\frac{1}{\lambda^{3}}g_{5/2}\left(  z\right)  -\frac
{2a}{\lambda^{4}}g_{3/2}^{2}\left(  z\right)  ,\nonumber\\
\frac{N}{V} &  =\frac{1}{\lambda^{3}}g_{3/2}\left(  z\right)  -\frac
{4a}{\lambda^{4}}g_{3/2}\left(  z\right)  g_{1/2}\left(  z\right)  ,\label{1}%
\end{align}
where $a$ is the scattering length, $\lambda=h/\sqrt{2\pi mkT}$ is the mean
thermal wavelength, and $g_{\sigma}\left(  z\right)  $ is the Bose-Einstein
integral. For obtaining the upper limit on the transition temperature for the
hard-sphere Bose gas, we need to find the virial expansion of Eq. (\ref{1}).
The second virial coefficient is \cite{Pathria}%
\[
b_{2}=\frac{1}{4\sqrt{2}}-2\frac{a}{\lambda}.
\]
Then the equation of state approximately reads%
\begin{equation}
\frac{PV}{NkT}=1-\left(  \frac{1}{4\sqrt{2}}-2\frac{a}{\lambda}\right)
\frac{N}{V}\lambda^{3}.\label{2}%
\end{equation}

By Eq. (\ref{2}) we can draw the $P-V$ diagram (Figure 1). Corresponding to
$\lambda>8\sqrt{2}a$ and $\lambda<8\sqrt{2}a$ there are two types of
isotherms: one type of the isotherms is monotonically decreasing, and the
other type of isotherms has a turning point --- a maximum point --- on the
curve. The isotherms that are not monotonic correspond to the transition.
Evidently, there is an unphysical part, corresponding to $dP/dV>0$, on each of
such isotherms. The correct result of this part should be isobaric. Between
this two type of isotherms, there is a critical isotherm which corresponds to
the upper limit on the transition temperature, $T_{u}$. Such an upper limit
can be determined by the relation $\lambda=8\sqrt{2}a$:%
\begin{equation}
T_{u}=\frac{h^{2}}{256\pi mk}\frac{1}{a^{2}}.\label{3}%
\end{equation}
When $T<T_{u}$, so long as the pressure is large enough, the transition may
occur. However, when $T>T_{u}$ (in this case the pressure $P$ is a
monotonically decreasing function of the volume $V$), the transition will not
occur regardless of how large may be the pressure. It should be emphasized
that in the case of the gas-liquid transition, instead of the "peak" which
corresponds to the transition there is a "kink" in the isotherm. This is
because when Bose-Einstein condensation occurs, the system is in the phase
which is a mixture of two phases all along the whole process of the transition
till the temperature tends to the absolute zero; in other words, in contrast
to the gas-liquid transition, there is no unmixed condensed phase in a Bose
gas due to the unachievable absolute zero.%
\begin{figure}[h]
\begin{center}
{\includegraphics[width=0.80\textwidth]{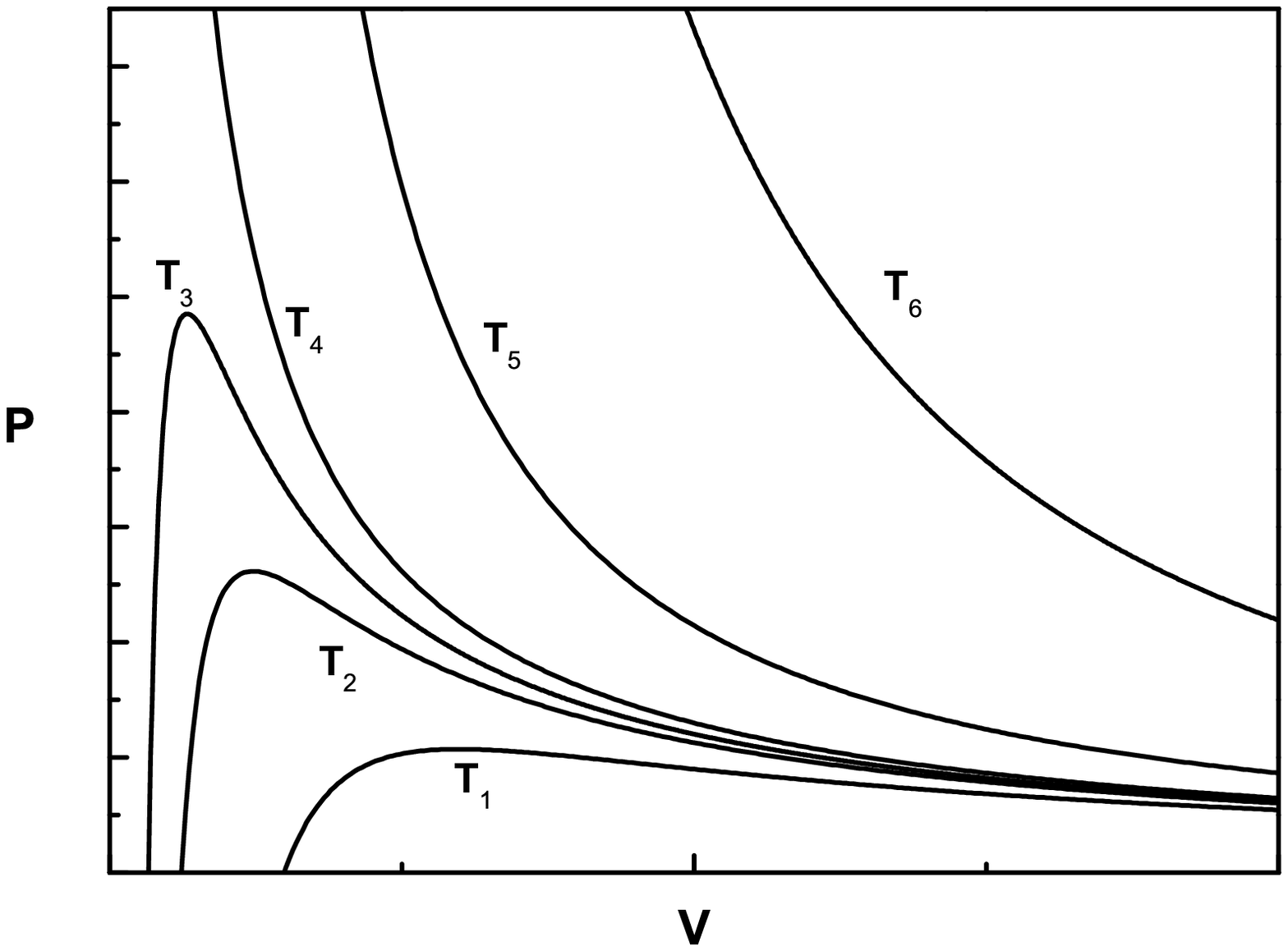}}
\end{center}
%\par
\begin{center}
{\small Figure 1. $P-V$ diagram of Eq. (\ref{2}).
$T_{1}<T_{2}<T_{3}<T_{u}$ and $T_{6}>T_{5}>T_{4}>T_{u}$.}
\end{center}
\end{figure}

The expression for the upper limit on the transition temperature, Eq.
(\ref{3}), shows that $T_{u}$ is inversely proportional to the square of the
scattering length, $a^{2}$. This implies that the stronger the interaction,
the lower the upper limit on the transition temperature. Especially, for ideal
gases, there are no interparticle interactions and so the scattering length
$a=0$; in this case $T_{u}\rightarrow\infty$, i.e., there is no restriction on
the transition temperature in ideal Bose gases. Such a conclusion agrees with
the above qualitative arguments.

Based on the above result we can estimate the upper limits on the transition
temperatures for some realistic Bose gases, $^{1}$H, $^{23}$Na, and $^{87}$Rb;
Bose-Einstein condensation in such gases have been observed \cite{Anderson}.
The spin-triplet scattering lengths of the atoms $^{1}$H, $^{23}$Na, and
$^{87}$Rb are $a^{H}=0.065nm$, $a^{Na}=2.8nm$, and $a^{Rb}=5.4nm$,
respectively \cite{Kleppner}. By Eq. (\ref{3}), we have
\begin{align*}
T_{u}^{H}  &  =5.6K,\\
T_{u}^{Na}  &  =0.13mK,\\
T_{u}^{Rb}  &  =9.3\mu K.
\end{align*}
Comparing these results to the transition temperatures of such atomic gases,
$T_{c}^{H}=50\mu K$, $T_{c}^{Na}=2\mu k$, and $T_{c}^{Rb}=0.67\mu K$
\cite{Kleppner}, we can see that the various $T_{u}$ obtained above are higher
than the transition temperatures. This agrees with the conclusion drawn above:
$T_{u}$ is the upper limit on $T_{c}$.

In conclusion, in this paper we show a difference between ideal and non-ideal
Bose gases: a non-ideal Bose gas has an upper limit on the transition
temperature, but an ideal one has none. If the temperature is higher than such
an upper limit, a non-ideal Bose gas cannot display Bose-Einstein
condensation. We estimate the upper limit on transition temperatures for some
realistic Bose gases and compare them to the transition temperatures measured
in some recent experiments.

Furthemore, the above result also implies that there may also exist an upper
limit on the transition temperature of superconductors, if superconduction can
be viewed as a kind of Bose-Einstein condensation.

\vskip0.5cm \textbf{Acknowledgement} We are very indebted to Dr G. Zeitrauman
for his encouragement. This work is supported in part by NSF of China, under
Project No.10605013 and No.10675088.

\end{document}